# Principles and Practices of Large-Scale Code Analysis at Ant Group: A Data- and Logic-Oriented Approach

Xiaoheng Xie
Ant Group
Shenzhen, China
xiexiemyself@gmail.com

Gang Fan
Ant Group
Shenzhen, China
fan.gang.cn@gmail.com

Xiaojun Lin
Ant Group
Shenzhen, China
linxiaojun.cn@outlook.com

Ang Zhou
Ant Group
Shenzhen, China
zhouang.za@antgroup.com

Shijie Li
Ant Group
Shenzhen, China
lishijie.lsj@antgroup.com

Xunjin Zheng
Ant Group
Shenzhen, China
zhengxunjin.zx@antgroup.com

Yinan Liang
Ant Group
Shenzhen, China
lyn249877@antgroup.com

Yu Zhang
Ant Group
Shenzhen, China
zy249870@antgroup.com

Na Yu
Ant Group
Shenzhen, China
yn304890@antgroup.com

Haokun Li
Ant Group
Shenzhen, China
lihaokun.lhk@antgroup.com

Xinyu Chen
Ant Group
Shenzhen, China
yancang.cxy@antgroup.com

Yingzhuang Chen
Ant Group
Shenzhen, China
chenyingzhuang.cyz@antgroup.com

Yi Zhen
Ant Group
Shenzhen, China
zhenyi.zhen@antgroup.com

Dejun Dong
Ant Group
Shenzhen, China
dejun.ddj@antgroup.com

Xianjin Fu
Ant Group
Shenzhen, China
xianjinfu@gmail.com

Jinzhou Su
Ant Group
Shenzhen, China
sujz12345@163.com

Fuxiong Pan
Ant Group
Shenzhen, China
fuxiong.pfx@antgroup.com

Pengshuai Luo
Ant Group
Shenzhen, China
luopengshuai.lps@antgroup.com

Youzheng Feng
Ant Group
Shenzhen, China
fengyouzheng@gmail.com

Ruoxiang Hu
Ant Group
Shenzhen, China
heat0415@gmail.com

Hanyang Guo
Ant Group
Shenzhen, China
guohy36@mail2.sysu.edu.cn

Jing Fan
Ant Group
Shenzhen, China
fredfanj@126.com

Xiao Xiao
Ant Group
Shenzhen, China
xx@antgroup.com

Peng Di
Ant Group
Shenzhen, China
dipeng.dp@antgroup.com

## Abstract

Large-scale software development requires dynamic and multifaceted static code analysis that extend beyond the capabilities of traditional tools. Existing tools like CodeQL lack cross-language analysis capability and are time-consuming and resource-intensive. We present CodeFuse-Query, a data system tailored for large-scale code analysis. First, CodeFuse-Query adopts a Logic Oriented Computation Design, employing Datalog with a two-tiered schema(COREF) to convert source code into data facts, and Gödel to express complex analysis tasks in logical terms. Furthermore,

[*]Corresponding authors: gfan@cse.ust.hk, dipeng.dp@antgroup.com.

CodeFuse-Query adopts Domain Optimized System Design. This approach optimizes resource utilization, prioritizes data reusability, applies incremental code extraction, and introduces tasks type characteristics specially for Code Change, underscoring its domain-optimized design.

We present empirical results demonstrating CodeFuse-Query's robustness, scalability, and efficiency in large-scale real-world scenarios at Ant Group, where it serves as a core static analysis infrastructure. Deployed in production environments, CodeFuse-Query processes up to 10 billion lines of code daily across over 300,000 distinct analysis tasks. CodeFuse-Query has been open-sourced[23].

## 1 Introduction

In the realm of software development at enterprise scale, IT organizations like Ant Group are confronted with the challenge of analyzing and maintaining exceptionally large volumes of code[26]. The inherent complexity of these vast codebases demands static code analysis solutions that surpass the capabilities of traditional tools such as the Clang Static Analyzer (CSA) and PMD[1, 5], which are effective for rule checking and style validation but often fall short in addressing the nuanced requirements of analyzing and managing large-scale code data.

Ant Group's diverse set of requirements for static code analysis includes not only the detection of bugs but also encompasses the analysis of code metrics across various programming languages to enhance R&D efficiency, monitoring compliance with complex regulatory algorithms, and the distillation of code features that are crucial for the development of robust defense mechanisms against network attacks. Addressing these multifaceted needs involves dealing with extensive code datasets that are characteristic of large codebases. The necessity to process and analyze such data efficiently-particularly in light of the computational constraints present in such large organizations-requires a system that integrates data-intensive processing capabilities with advanced program analysis algorithms.

Nowadays, there are some tools for large scale code analysis. CodeQL [2] is one of the most popular static analysis tools provided by GitHub to help developers analyze, review, and improve the quality of their code. It automates the identification of potential security vulnerabilities, logic errors, and performance issues in code, and is a code-oriented query language that allows users to write queries to discover potential problems. It works in a database-based manner, allowing efficient analysis of large code bases. We select it as our baseline. However,

To address this, we have designed and built CodeFuse-Query within Ant Group. CodeFuse-Query adopts a data-centric approach to code analysis, transforming analyzing tasks into data computation tasks. CodeFuse-Query currently supports more than three hundred unique static analysis tasks and is used in more than thirty different scenarios. Additionally, CodeFuse-Query is compatible with nine different programming and configuration languages. The system processes an approximate workload of 300,000 tasks daily and scans in excess of ten billion lines of code each day.

Central to CodeFuse-Query are two strategic design principles that direct its technical development: The first strategy, *Domain Optimized System Design*, involves creating solutions meticulously suited to the requirements of static code analysis. This includes the development of task types for "Code Change" analysis that cater to the incremental development of codebases, as well as the implementation of a resource-aware scheduling strategy to effectively manage computing resources while giving priority to data reusability. Additionally, the system adopts an incremental code extraction strategy to efficiently process the continual changes within large, dynamic codebases, providing accurate and up-to-date analysis without excessive computational use.

The second strategy, *Logic Oriented Computation Design*, treats code analysis as a logical inference process. CodeFuse-Query utilizes Datalog to enable complex reasoning over codebases. The core of this approach is the Code Interpretation Format (COREF) schema, which transforms source code into a set of data facts. Complementing this is Gödel, a language designed to exploit Datalog's declarative properties, allowing for the formulation of intricate code analysis tasks as logical queries. The effectiveness of CodeFuse-Query within this design category is evidenced by its successful application to various programming and configuration languages. The efficacy of CodeFuse-Query is supported by empirical data through its application to nine different programming and configuration languages.

CodeFuse-Query introduces a new approach to static analysis, meeting domain-specific requirements while accommodating large-scale software development dynamics. CodeFuse-Query has supported 9 programming languages as Java, Js/Ts, XML, Python. This paper suggests that CodeFuse-Query, with its approach to static code analysis, offers a potential path for managing the escalating complexity and diversity of large-scale software development. The paper's primary contributions are:

(1) **Novel Query Approach with CodeFuse-Query:** The paper presents CodeFuse-Query, an open-sourced data-centric system that analyzes code with its domain-optimized and logic-oriented design. Using Datalog, CodeFuse-Query transforms source code into data facts through a two-tiered schema, COREF, and a user-friendly language, Gödel, which simplifies complex tasks.
(2) **Empirical Evidence for CodeFuse-Query's Design:** The design of CodeFuse-Query is supported by empirical evidence. This paper presents tests and evaluations that provide insights into CodeFuse-Query's robustness, scalability, and efficiency, validating its effectiveness in query-like large-scale static analysis.

(3) **Real-world Impact of CodeFuse-Query:** This paper outlines practical applications of CodeFuse-Query. We presents use cases, such as generating Business Intelligence (BI) from code data and analyzing training data for large language models (LLMs) [25], to show the practicality of CodeFuse-Query. We also document how CodeFuse-Query can be integrated with nine different programming and configuration languages. CodeFuse-Query has been open-sourced ??.

## 2 Background and Challenge

The CodeFuse-Query system is designed to address numerous challenges encountered in real-world scenarios and applications.

**Problem Complexity and Dynamism (Complexity Challenge)**: CodeFuse-Query must handle a wide range of tasks with diverse requirements. These tasks include ad-hoc BI reports on software repositories, strict code gate checks for programming standards, comprehensive security vulnerability assessments, and fully automated, precise incremental code change impact analyses. Each task varies in domain, performance needs, and automation level, contributing to the complexity.

**Diversity of Analysis Targets (Diversity Challenge)**: The system needs to support a broad spectrum of products-backend services, frontend components, mobile clients, and mini apps-developed using various programming languages and frameworks. Each may utilize specific libraries and middleware solutions for various functions. The core challenge is to provide accurate insights across this heterogeneous environment, regardless of the languages or frameworks used.

**Scalability and Speed (Efficiency Challenge)**: With billions of lines of existing code and over 100,000 new lines added daily at Ant Group, CodeFuse-Query must efficiently scale to analyze large codebases. It needs to manage vast amounts of code and integrate new code promptly to maintain up-to-date analyses, ensuring relevance in a fast-paced development environment.

**Resource Constraints and Variability (Resource Challenge)**: Static analysis is computationally intensive due to path explosion, and ideal accuracy requires significant resources. However, in practice, resources are limited and shared among projects with varying priorities. CodeFuse-Query must optimize resource usage and perform consistently despite resource fluctuations, balancing accuracy and efficiency within real-world constraints.

**Evolutionary Maintenance (Maintenance Challenge)**: As software paradigms evolve and codebases grow in complexity, CodeFuse-Query must remain easily updatable, extensible, and maintainable. This includes accommodating new languages, frameworks, and coding practices, updating algorithms for new programming constructs, and fixing bugs. Limited resources and the potential accumulation of technical debt [53] further emphasize the need for an adaptable and efficient platform design.

These challenges span technical, operational, and strategic aspects, all critical for managing a large-scale static code analysis platform effectively. Addressing them ensures that CodeFuse-Query remains robust, accurate, and efficient in supporting diverse and evolving real-world software development needs.

## 3 Domain Optimized System Design

The primary design objective of CodeFuse-Query is to create a code analysis platform with a focus on scalability, performance, reusability, and extensibility. In the following sections, we will lay out our design principles in more detail, describe the features of CodeFuse-Query, and discuss our system architecture.

### 3.1 Design Principles

**Domain-Integrated Design**: To tackle the *Complexity and Diversity Challenges*, we integrate domain-specific features into the system design, focusing on the unique requirements of code analysis tasks. By partitioning analysis into Extraction and Query phases, we balance detail richness with storage efficiency.

**Maximal Data Reusability**: Addressing the *Efficiency Challenge*, we enhance data reuse across the processing chain. Caching at the Proxy layer, incremental extraction using the Facts DB, and in-memory data utilization at the Analysis Node reduce redundant computations and improve performance.

**Computation Optimization**: To meet the *Efficiency Challenge*, we implement strategies like fixed-point computation and an optimized extract-query design, reducing computational demands and system overhead during query execution.

**Resilience Through Redundancy**: Confronting the *Maintenance Challenge*, we design for resilience by incorporating redundancy. Anticipating potential failures, we employ internal retries, task rescheduling, and mitigate single points of failure to ensure uninterrupted operation.

**Dynamic Resource Allocation**: Addressing the *Resource and Efficiency Challenges*, we dynamically allocate computational resources in response to workload fluctuations. By decoupling system components and separating extraction from querying, we optimize resource utilization and maintain high performance.

These principles ensure that CodeFuse-Query remains robust, efficient, and adaptable, effectively supporting large-scale static code analysis across diverse and evolving real-world software development environments.

### 3.2 System Interface

CodeFuse-Query leverages the query language Gödel (Definition 3.2), enabling developers to specify complex code analysis requirements precisely. The language supports operators for pattern matching, code structure, and code similarity, allowing detailed information retrieval criteria.

The system offers functionalities categorized into **Extraction**, **Analysis**, and **Scanning** operations: *extract()*: Constructs the code data model, outputting the facts database formatted as COREF (Definition 3.1). *analysis()*: Performs advanced code analysis based on user-provided queries (Section 3.2.1), a specified code file or repository URL, and its version identifier. For new projects, analysis begins with a full extraction to build the facts database. *scan()*: Detects and extracts code features and structures matching defined patterns, applicable to single code files or entire repositories.

When building the facts database, CodeFuse-Query checks for existing snapshots matching the version and incrementally constructs model parts by comparing file differences and code dependencies against a baseline snapshot. Each version extraction creates a new snapshot. Different repository versions can be extracted concurrently, but within the same repository and data version, only one extraction task exists, ensured by database row-level locking.

*3.2.1 Terminology.* As mention above, we partition static analysis into Extraction and Querying. The "Extraction Result" comprises information extracted from source code and transformed into a structured data model called COREF (Definition 3.1). To facilitate analysis, we developed *Gödel* (Definition 3.2), a simple and accessible programming language that allows developers to write queries for detailed examination of the modeled data.

**Definition 3.1 (COREF)** A comprehensive representation scheme for code data, employing a two-tiered data modeling approach. This methodology represents the complex structure and semantics of code, facilitating effective analysis (Section 4.1).

**Definition 3.2 (Gödel)** A Domain-Specific Language (DSL) tailored for defining and executing code analysis. Gödel also serves as a foundational computational engine. Built on the principles of Datalog, it adheres to the same computational model as a typical Datalog program (Section 4.3).

*3.2.2 Analysis Task Structure.* Our system integrates various task types for code analysis, each tailored to specific needs and contexts.

The *Full Repository Analysis (FRA)* examines the entire code repository as a single unit, considering file interconnections when building the code database. FRA is foundational for other tasks and is essential for large-scale repository analysis, such as extracting metrics or performing ad hoc scans.

Code changes are crucial in software development, affecting security, risk management, and productivity. Our system accommodates this with two code change analysis tasks to enhance efficiency and meet time constraints: *Incremental Full Repository Analysis (IFRA)*: Updates the database incrementally when code changes occur, enabling faster and more efficient full repository analysis. *Delta Code Analysis (DCA)*: Focuses only on the altered files instead of the entire repository. For example, when identifying functions impacted by a code change, DCA analyzes functions in modified files, while IFRA considers all functions in the repository.

## 3.3 System Architecture

CodeFuse-Query utilizes a service-oriented architecture, achieving clean decoupling among system components. Figure 1 shows that CodeFuse-Query is composed of four layers: the access layer, coordinator layer, worker layer, and storage layer.

The *Access layer* uses stateless proxies as user endpoints, operating concurrently to route client requests to processing units and aggregate results. These proxies cache metadata for task verification (e.g., confirming code file existence) and cache results to reduce redundant analyses, beneficial in continuous integration and deployment. This approach allows early rejection of failed verifications, reduces overhead from repeated analyses, and decreases request latency by minimizing routing hops.

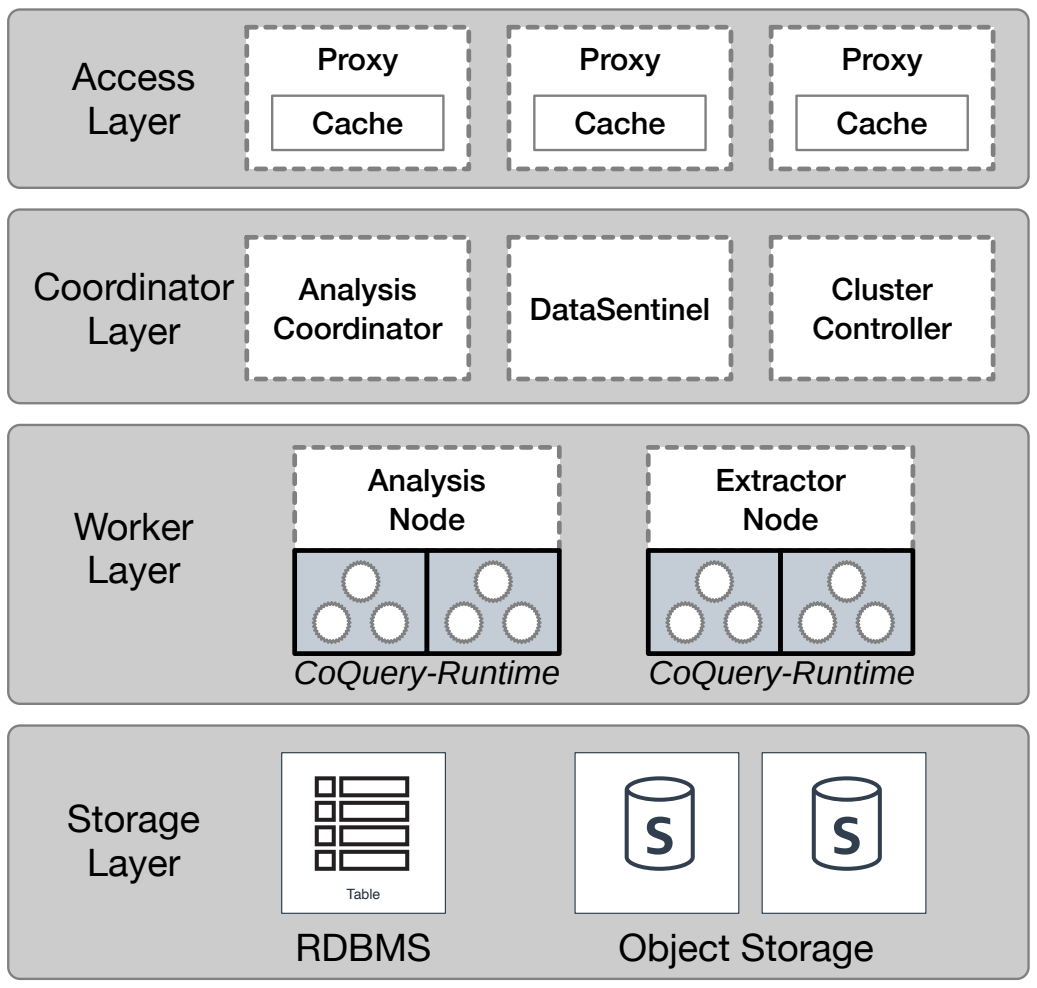

**Figure 1: The Architecture of CodeFuse-Query**

The *Coordinator layer* manages system status, maintains metadata, and oversees task processing. This layer houses three components, each with specific responsibilities:

(1) *Analysis Coordinator*: Supervises analysis nodes, manages load balancing, task splitting and aggregation, and monitors execution status. It distributes tasks across nodes to balance workload and traces query execution plans to include dependencies, checking for unchanged files to optimize task assignment. If files are unchanged, it aligns results from DCA, IFRA, and FRA, transforming FRA tasks into IFRA tasks to reuse analyzed data and prevent redundant processing. An estimation component calculates expected computational costs based on factors like programming language, lines of code, and repository size, optimizing resource allocation for large-scale tasks.
(2) *Data Sentinel*: Manages code facts DB metadata (versions, repository addresses, commit IDs), schedules code modeling and feature extraction, and oversees task status. It ensures balanced data distribution across clusters, especially for frequently accessed data, and synchronizes Extractor nodes. For new repositories, it performs a full COREF construction; for code modifications, it updates the database incrementally, reducing extraction node load, extraction time, and disk I/O.
(3) *Cluster Monitor*: Monitors cluster status, tracking metrics like memory and CPU usage, and dynamically scales clusters during high workloads. Components operate as equal instances, with the Cluster Monitor regularly checking their status. Replicas take over failed components, ensuring fault tolerance, load balancing, and scalability.

The *Worker layer* executes computational tasks using stateless worker nodes that retrieve read-only data copies without dependencies. This design allows elastic scaling under high computational loads and enables resource isolation for varying QoS requirements. We have two types of workers: extractor and analysis nodes. Both are implemented consistently, possessing capabilities for modeling and analysis, allowing seamless transformation between them.

The *Storage layer* in CodeFuse-Query persists system status, metadata, and the code facts database. We use a Relational Database Management System (RDBMS) for status and metadata, and an object storage service for the large-volume code facts DB stored in SQLITE format [38]. The storage layer supports real-time analysis by distributing frequently accessed data to worker nodes, reducing reads from object storage. Additionally, we deploy computations directly on the nodes storing the data, following a Data-Driven Computing approach.

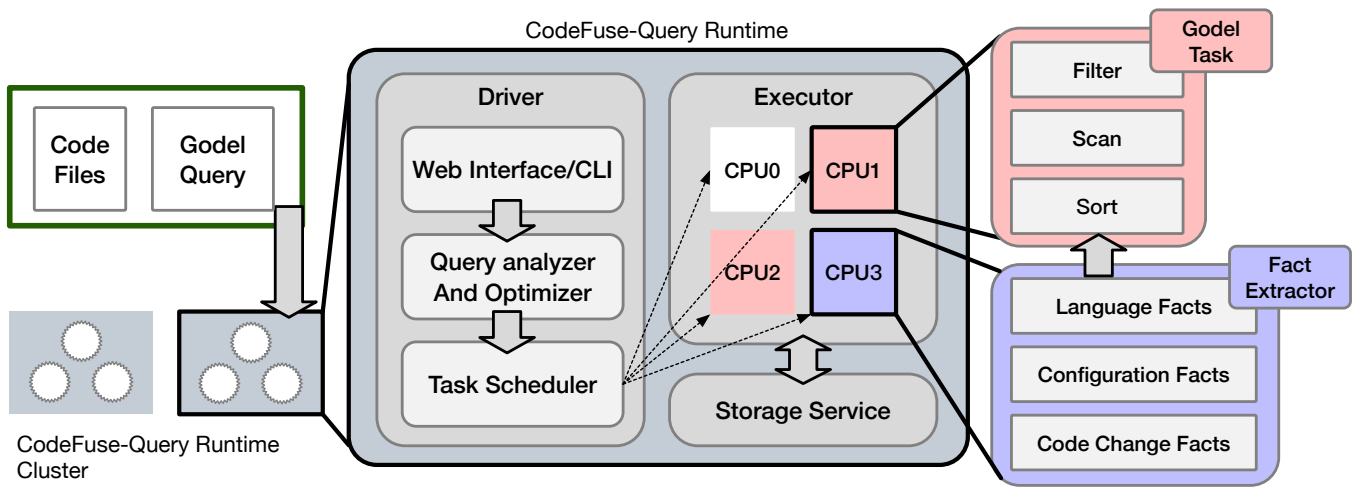

**Figure 2: The worker layer runtime architecture**

## 3.4 The CodeFuse-Query Runtime

The CodeFuse-Query Runtime component, shown in Figure 2, manages query execution, code modeling, and feature extraction. Applications sent to the Runtime are denoted as jobs, each subdivided into stages. A stage represents a segment of a job, such as a reading stage for file input, culminating in either a data exchange or the final result. Stages in CodeFuse-Query Runtime are block-oriented, with the next stage starting after the preceding stage concludes, allowing fault tolerance via stage replaying. The Runtime supports multiple concurrent jobs.

The Runtime uses a Driver for scheduling, query optimization, and data alignment. A Driver controls multiple Executors, each handling data scanning, manipulation, and result generation. Executors perform various operations such as Gödel script computation, source code modeling, feature extraction, and User-Defined Functions (UDFs). Executors are multi-threaded, incorporate task scheduling and thread pooling, facilitating parallel execution of independent tasks. Tasks that need exclusive resources are queued.

The structured approach of the CodeFuse-Query Runtime shapes the system's computational design for code analysis. Through its strategic stages and concurrent job handling, CodeFuse-Query enables efficient, fault-tolerant processing. The next section will detail these design elements, outlining the computational process that bolsters our large-scale code analysis system.

# 4 Logic-Oriented Computation Design

The heart of *CodeFuse-Query*'s runtime engine is its logic-oriented computational design, which is crucial for efficiently analyzing vast volumes of source code. This section details the design elements for large-scale code analysis.

Figure 3 provides a comprehensive overview of the computational process, which begins with the interpretation of user queries into a Gödel program and the extraction of source code facts into a database. The runtime of our system then executes the program to produce the analysis results. The computational process comprises the following steps:

(1) Users write queries in Gödel. These queries are interpreted by the system and turned into a comprehensive Gödel program. This program includes the analysis algorithm library, user query-derived custom code, and the Tier-2 COREF model, also in Gödel.
(2) After the program code is prepared, our system generates an execution plan from the Gödel program. This plan encapsulates the computational relationships among facts, leading to the final facts. We then apply optimization techniques to enhance the plan's efficiency. Not all facts require reasoning, offering significant optimization potential. For instance, we can reduce a plan from 2832 computations to just 17.
(3) After optimization, we compile the execution plan into an executable binary. By integrating the extracted facts from the source, we engage in a logical reasoning process to execute it, culminating in the production of results.

Subsequent subsections delve deeper into the design of the underlying computational model, datalog, the two-tiered data modeling approach, and the DSL, Gödel, used to frame the computation.

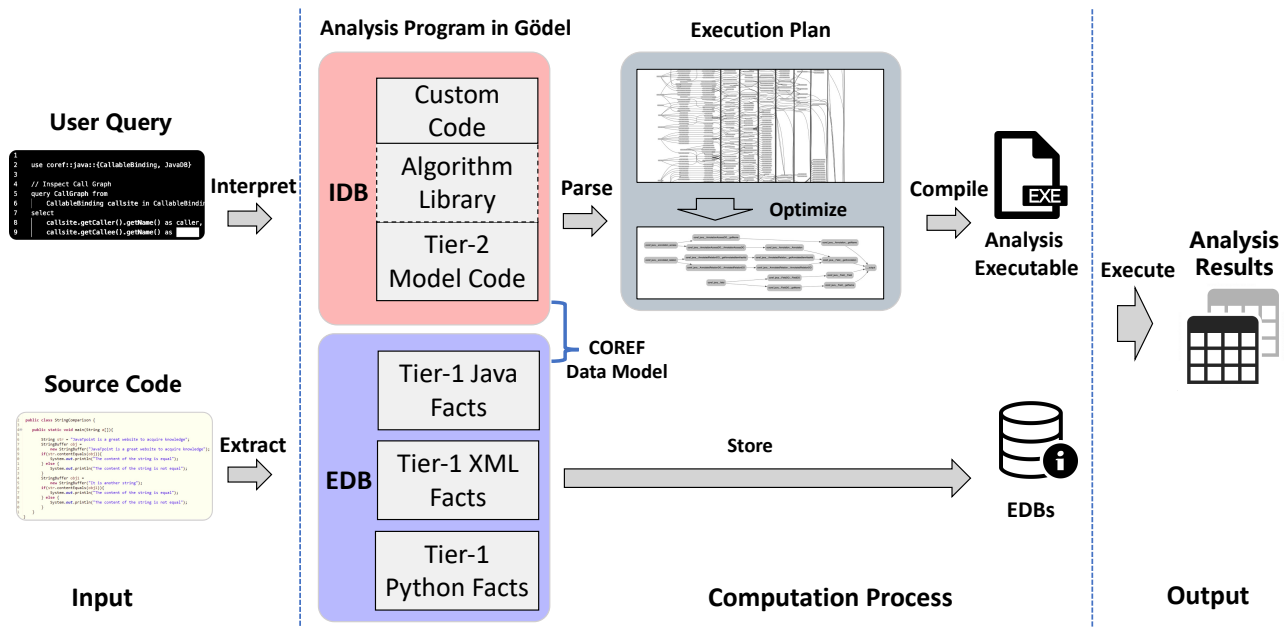

**Figure 3: The Computation Design**

## 4.1 Underlying Computation Model: Datalog

Gödel, a programming language for formulating the computation of CodeFuse-Query, is fundamentally built upon Datalog[20], a declarative logic language[36] that abstracts complex computations, permitting users to specify their requirements without fretting over execution minutiae. Datalog's human-friendly syntax mirrors pseudo code, facilitating easy writing and maintenance. Moreover, its capabilities in handling complex queries, recursion, and logical reasoning make it suitable for static code analysis.

Consider an illustrative Datalog program that computes ancestor-descendant relationships of classes, using predicates parent and ancestorclass. The rules are defined as follows:

```
parent(a, b) :- class(a),
                extends(a, b),
                class(b).
ancestorclass(a, b) :- parent(a, b).
ancestorclass(a, c) :- parent(a, b),
                       ancestorclass(b, c).
```

Datalog's computation model is a specialized relational algebra variant, enabling logical computations among facts and supporting recursive computation[16], a requisite for many static code analysis tasks[15].

Datalog's computational prowess, harnessed by GÖDEL, is rooted in several key attributes. It leverages optimization algorithms to streamline execution, ensures program termination to prevent infinite loops, operates set-oriented computations for efficiency, and handles complex queries and recursive computations proficiently. In practice, we use Soufflé[40] as the Datalog engine for *GÖDEL*. Other available engines include $\mu$Z[39] and DDlog[44]. While Datalog serves as the computational core of *GÖDEL*, our DSL extends it with higher-level features. Further details on the enhancements provided by GÖDEL will be explored in Section 4.3.

## 4.2 Code Modeling and Extraction

Our system is underpinned by a data-centric approach, with the comprehensive schema, COREF, playing a critical role in the modeling of source code. COREF encapsulates both the syntactic structure and the semantic information intrinsic to the code.

COREF provides a suite of common code facts for analysis, including elements drawn from standard source code representations such as Abstract Syntax Trees (AST), Control Flow Graphs (CFG), Abstract Semantic Graphs (ASG), and Program Dependency Graphs (PDG). These elements deliver insights into syntax, control flow, semantics, and dependencies within the code, thereby establishing a robust foundation of standard code facts for users to analyze and query.

This proactive approach to data retention allows us to confront two major challenges in code analysis: the *Complexity Challenge*, and the *Diversity Challenge*. By extracting the necessary data on-demand, COREF supports a broad spectrum of analyses. Despite the vast array of information in the source code, storing all data in a database is not practical. However, our "Two-Tiered Modeling" technique circumvents this issue, successfully storing all relevant code information at a reasonable cost.

*4.2.1 Two-tiered Data Modelling.* Figure 3 depicts our data modeling process. The initial step is the extraction of source code facts, stored in a local database, represented by our Tier 1 model extraction, comprising roughly a hundred tables per programming language.

The Tier 1 model is designed primarily to store facts in the most compact form, not for direct user querying. This focus on efficiency prevents duplicate information storage and enables recomputation of other necessary facts. Thus, the database is designed for efficient persistent storage. For example, facts for analyzing a typical Java repository with about 1M lines of code can be stored in just around 720MB in the database. For comparison, precomputing and expanding all the facts for a typical Java repository would take around four hours and 3.1GB of disk space. The Tier 2 model is where user interaction mainly occurs, providing facts ready for querying. Typical facts, like a class's qualified name, are calculated by analyzing multiple tables from the Tier 1 facts. The synthesis of information occurs dynamically at execution time, with Tier 2 facts accessible as if precomputed. Essentially, Tier 2 facts are computed as needed, providing on-demand access to complex data relationships. In Datalog terms, our Tier 1 model is similar to the "Extensional Database" or EDB[50], containing ground facts available before computation. The Tier 2 schema, in GÖDEL, is the "Intensional Database" or IDB, consisting of derived facts, emerging from the reasoning process applied to EDB facts.

In conclusion, our two-tiered approach with user-specific queries allows us to interconnect all data and encapsulate all static analysis tasks as Datalog computation tasks. This method provides an efficient means for managing and analyzing source code information. We have support nine program languages, developed mature code models for Java, XML, and JavaScript/TypeScript, and beta-level models for Objective-C, Go, Python, Swift, SQL, and others.

*4.2.2 Language-Specific Fact Extraction.* Our primary strategy for constructing language facts is *language-specific fact extraction*. We develop unique extractors and apply distinct COREF schemas for each language, retaining crucial grammar details vital for many applications. This contrasts with Universal Abstract Syntax Trees (UASTs) and Intermediate Representations (IRs), which aim for a unified representation but may lose language-specific nuances. Building parsers in their native languages ensures optimal compatibility and ease. With the use of Language-Specific Fact Extraction, CODEFUSE-QUERY can achieve cross-language analysis.

While this approach requires significant engineering effort to create extractors for each language, it results in a scalable and standardized process applicable to multiple languages. Importantly, our parsers operate *independently of build processes* like Maven for Java or Bazel for C++. This strategic choice addresses the *Maintenance Challenge* by enhancing system maintainability and simplifying debugging, as build processes can be unreliable and brittle in large organizations. This method differs from other static analysis tools like CodeQL [2, 14], which depend on build systems to capture accurate source code and compilation information. Although our approach may lead to some loss of information—especially in languages with preprocessing stages like C—we mitigate this with fuzzy parsing techniques.

Crucially, while our modeling and extraction are language-specific, our queries are not. This design allows us to seamlessly analyze facts across different languages, effectively addressing the *Diversity Challenge*.

## 4.3 Formulating the Computation in GÖDEL

We have developed GÖDEL, a Domain-Specific Language (DSL), specifically tailored for articulating code analysis computations. GÖDEL's primary design goals are twofold: 1) to provide a user-friendly interface enabling users to query and compute facts easily, addressing *Complexity and Diversity Challenge*) to implement an efficient and maintainable approach capable of handling the high volume and continue changing of tier-2 modeling. Ultimately, a GÖDEL program will be compiled into a Datalog program, which is then transformed into an execution plan and executable. While this paper does not explore the comprehensive design and implementation details of GÖDEL, it underscores several key design choices that exemplify the vital role GÖDEL plays in large-scale code analysis.

*4.3.1 A SQL-like Interface Coupled with Rust-like Constructs.* In GÖDEL's design, we merge two distinctive strategies. The first caters to common use cases such as querying facts with light analysis and manipulation, for which we offer a SQL-like query interface to minimize the learning curve. As shown in Listing 1, we provide a SQL-like style example to find the text info and location info of

**Listing 1: A Query Example**

```
use coref::java::{Import, JavaDB}

fn db() -> JavaDB {
    return JavaDB::load("coref_java_src.db")
}

query importStatement from
    i in Import(db())
where
    i.getName().contains("springframework")
select
    i.getName() as name,
    i.getLocation().getFile().getRelativePath() as filepath
```

any import statements with the name "springframework" in the codebase. However, SQL is known to be challenging to maintain when the analysis method is highly complex[37, 43]. For instance, we have queries that span over 700 lines of script code. In such scenarios, we shift the language to a Rust-like code that includes high-level features such as Structs, Functions, and Types, which are leveraged to write intricate algorithms. Intriguingly, in Gödel's DSL design, these two methods, despite their differences, can coexist within one script. This hybrid approach accommodates a broad spectrum of tasks, from lightweight operations like querying a class name to intricate algorithms such as customized dataflow analysis, effectively addressing both *the Complexity and Diversity Challenge* as well as *the Maintenance Challenge*.

*4.3.2 Rich and High-level Program Features.* Gödel provides a rich set of high-level features, enhancing the process of complex code analysis. Rather than just being a Datalog reiteration, it is a specialized language for code analysis and modeling, combining Datalog's computational power with a more expressive and user-friendly interface. Gödel excels in abstracting complex logic into understandable structures, simplifying the expression of intricate algorithms and addressing the increasing complexity challenge often encountered in traditional Datalog. Unlike traditional Datalog, which offers limited data types, Gödel has a robust type system that allows for complex data type definitions and enforces strict type checks, enhancing expressiveness and data integrity. Gödel's modularity facilitates code reuse, readability, and maintenance by partitioning programs into distinct modules. This stands in contrast to the monolithic structure common in traditional Datalog programs. Additionally, Gödel includes control structures such as recursion, conditional statements, and looping constructs. These features, which can be transformed into corresponding Datalog semantics, expand Gödel's versatility while preserving Datalog's computational efficiency. Gödel also incorporates UDFs and foreign functions, making it a Turing-complete language and enhancing its adaptability and flexibility for specific computational tasks. In practical applications, using Gödel can lead to a minimum of 50% reduction in source code lines compared to Soufflé, resulting in code that is not only more maintainable but also easier to comprehend.

*4.3.3 Utilizing Common Algorithms via Gödel Library.* Gödel integrates a robust Common Algorithm Library, housing various predefined and user-defined modules. This code repository encourages reusability and enhances computational efficiency. The library contains frequently utilized static analysis algorithms like control flow analysis, escape analysis, dataflow analysis, and live variable analysis. Users can readily use these predefined algorithms, bypassing the need for reimplementation. The library also supports user-defined modules, facilitating customization per project needs. For instance, a user might create a custom taint analysis algorithm, package it as a module, and reuse it in multiple projects. The Gödel compiler is optimized for this modularity, effectively executing these libraries and ensuring efficient translation of high-level Gödel code into Datalog programs.

## 4.4 Optimizations and Opportunities

The resource-intensive nature of static code analysis necessitates strategic optimization to conserve CPU and memory utilization. Gödel, designed with a focus on efficiency, introduces dual-level optimization techniques: one intrinsic to the Gödel platform, and the other targeting the Datalog computation process. Both these strategies jointly address *Efficiency Challenge*, critical to large-scale code analysis. Gödel programs are convertible into execution plans, which distill complex computations into a streamlined process, providing a platform for optimization. Application of sophisticated optimization algorithms[15] can drastically decrease the complexity of standard queries. As illustrated in Figure 3, the implementation of these algorithms can reduce an analysis from 2832 computation nodes to a mere 17, without compromising the results.

Beyond this program-level optimization, Gödel's architecture also enables system-level enhancements. The system can cache "hot" facts—data points that are frequently computed and accessed, thereby curbing the redundant computation of repetitive facts. Moreover, Gödel's design allows for proactive computation, where the system can anticipate and calculate certain facts before a query is initiated, based on usage pattern analysis. This preemptive approach bolsters the system's overall efficiency, ensuring it can adeptly handle large-scale code analysis tasks with optimized resource utilization.

# 5 Evaluation

We evaluate our system's performance by analyzing computing efficiency against code volume and query execution across varying database sizes and complexities. Our evaluation focuses on three areas: comparative testing with CodeQL [2], task scheduling efficiency, and system reusability.

## 5.1 Experimental Setup

**Datasets**. For our experiments, we utilized the following codebases:

- Ant Group codebases: A selection of 4,000 codebases, ranging from 100,000 to 2,000,000 lines of Java language.
- Open-source codebases: A random sample of 50 open-source repositories from GitHub of Java and Python languages.

**Setup**. Evaluations were conducted by two distinct configurations:

- A single machine (MacBook Air with Apple M1, 16GB memory, and an 8-core CPU) was used to assess the performance of CodeFuse-Query Runtime and CodeQL.

- A cluster environment (comprising 300 machines, each with 8GB RAM and a 4-core CPU) was employed to evaluate IFRA, DCA, and HDT tasks.

## 5.2 Comparison of CodeFuse-Query Runtime and CodeQL

**Table 1: Comparison of Repository Analysis Success Rates**

| Dataset | CodeFuse-Query Runtime | | CodeQL | |
|---|---|---|---|---|
| | Suc./Total | Rate (%) | Suc./Total | Rate (%) |
| Java | 35/35 | 100 | 10/35 | 28.5 |
| Python | 15/15 | 100 | 15/15 | 100 |

**Extraction Performance.** CodeQL supports three build modes: none, for interpreted languages, C#, and Java; autobuild, for compiled languages; and manual, where build steps are defined in the workflow. In contrast, CodeFuse-Query Runtime is designed to extract code without the need for a build process for all languages. As shown in Table 1, we evaluated 50 open-source codebases to compare the extraction success rates of CodeFuse-Query Runtime and CodeQL. For Java language, CodeFuse-Query Runtime achieved a 100% success rate, while CodeQL succeeded in analyzing only 10 out of the 35 codebases by using autobuild mode, resulting in a 28.5% success rate. The lower success rate of CodeQL is attributed to its reliance on the build process, which may fail due to various reasons such as missing dependencies or configuration issues. For Python, both tools adopt a build-independent approach and equally achieved a 100% success rate. We also evaluated code extraction performance in build-independent mode (Table 2). For Java projects averaging 75.57 KLOC, CodeFuse-Query Runtime completed extraction in 33.47 seconds on average, whereas CodeQL required 912.31 seconds due to dependency preparation. For Python projects averaging 24.47 KLOC, CodeFuse-Query Runtime had an extraction time of 6.41 seconds, compared to CodeQL's 15.37 seconds. Further comparison on 10 Java repositories in build mode (Table 3) showed that CodeQL's total processing time was approximately 5.66 times longer than that of CodeFuse-Query Runtime.

The results show that CodeFuse-Query Runtime's build-independent design significantly enhances extraction success rate and performance, especially for compiled languages. By eliminating the need for the build process, it efficiently supports large-scale analysis of complex projects.

**Table 2: Comparison of Code extraction Performance Over 50 Open-source repositories with build-independent mode.**

| Dataset | LOC | CodeFuse-Query Runtime | | CodeQL | |
|---|---|---|---|---|---|
| | | Times(s) | Mem(MB) | Times(s) | Mem(MB) |
| Java | 75k | 33.47 | 4193.56 | 912.31 | 2208.51 |
| Python | 24k | 6.41 | 300.99 | 15.37 | 1352.28 |

**Evaluation of Query Performance.** We evaluated the query performance of both CodeFuse-Query Runtime and CodeQL by executing 15 representative queries. The results are summarized in Table 4. We observe that CodeFuse-Query Runtime generally outperforms CodeQL in both query execution time and memory usage, especially for Python repositories. For Java datasets, CodeFuse-Query Runtime was approximately 1.23 times faster than CodeQL, with comparable memory consumption. In the case of Python datasets, the performance gap is more pronounced. CodeFuse-Query Runtime is approximately 1.72 times faster than CodeQL and CodeQL used about 9.68 times more memory. The superior performance of CodeFuse-Query Runtime can be attributed to its optimized execution plan, which loads only essential information into memory during query processing. Furthermore, the query results from CodeFuse-Query Runtime were consistent with those obtained from CodeQL, which indicates that despite the differences in execution time and memory usage, both tools are effective in detecting the intended code patterns and potential issues.

**Table 3: Comparison of Code extraction Performance Over 10 Java Open-source repositories with Build mode.**

| | CodeFuse-Query Runtime | | CodeQL | |
|---|---|---|---|---|
| | Extraction | Build | Extraction | Build |
| Time(s) | 221.4 | - | 633.7 | 598.2 |

**Table 4: Comparison of Querying Performance.**

| Dataset | CodeFuse-Query Runtime | | CodeQL | |
|---|---|---|---|---|
| | Time(s) | Mem(MB) | Time(s) | Mem(MB) |
| Java | 13.6 | 1159.3 | 16.7 | 1207.8 |
| Python | 9.2 | 164.5 | 15.8 | 1592.8 |

## 5.3 Evaluating the Efficiency Designs

Our evaluation aims to showcase the efficiency and adaptability of our system in managing extensive and varied modifications across large codebases, to assess the performance enhancements brought about by three specific mechanisms: IFRA Task, DCA Task, and Dedicated Long-run Queueing.

For the evaluation of IFRA and DCA Task mechanisms, we record the true code changes in the codebases and isolate the system queue time and network latency to assess the performance of these two mechanisms.

On the other hand, to evaluate the Dedicated Long-run Queueing mechanism, we record tasks from the live system and replay them multiple times under various conditions. This approach allows us to observe the scheduling mechanism's effectiveness in a real-world setting. By replaying the same tasks multiple times, we can assess the consistency of the system's performance and its ability to optimize overall end-to-end time.

**The Efficacy of the IFRA task design.** We compare the performance of the IFRA task design with the traditional FRA task design by using 8 large-scale open-source codebases [28–35]. For each codebase, we evaluated both methods on a subset of 10 commits, measuring execution time and maximum memory usage. Table 6 presents the comparison results. The IFRA task achieved a significant reduction in extraction time, decreasing from an average of 267.01 seconds (FRA) to 38.68 seconds-an 83.18% reduction. Memory usage was also substantially lowered, with an average reduction of 84.8% . These reductions were consistent across all projects, regardless of codebase size. By significantly reducing both extraction time and memory consumption, IFRA is particularly well-suited for large-scale code analysis, especially in environments with limited system resources.

**The Efficacy of the DCA task design.** To assess the efficiency of the DCA task methodology, we conducted a comparative analysis with the FRA task implementation across the same set of 140 tasks by analysing the selecting 22 open-source repositories. As depicted in Table 5, The results indicate that on average, a DCA task required 3.67% of the analysis time compared to a FRA task, suggesting that the DCA task approach may be more time-efficient for processing code modifications.

Furthermore, the role of the scheduler in this process is an important factor to consider. Upon determining that a task can be downsized to a DCA task, the scheduler not only conserves substantial system resources but also significantly enhances performance. This decision-making capability of the scheduler appears to contribute to the system's overall performance and efficiency.

**The Efficacy for handling High-Duration Tasks.** We conducted an experiment analyzing gateway interfaces in a dataset of 4,000 repositories to investigate the effect of task execution time on the overall workflow and to explore optimization strategies. This experiment utilized actual tasks from a production environment to enhance the relevance and applicability of the results.

In our workflow, we set a 3,600-second limit for each task's execution time. Tasks exceeding this limit were termed "High-Duration Tasks" (HDTs). Initially, tasks dispatched to CodeFuse-Query without specific strategies resulted in a total execution time of 10,488 seconds for 4,000 tasks, including failed tasks and those that hit the time limit, primarily due to 11 HDTs within the set (see Table 7 for details).

To optimize this, our Analysis Coordinator uses a strategy where HDTs are identified and directed to a separate queue processed by higher-capability machines. This strategic resource allocation resulted in a reduction of the total execution time to 4,006 seconds, corresponding to an efficiency improvement of approximately 62%. Additionally, this strategy ensured a 100% success rate, demonstrating its effectiveness in managing HDTs.

The outcomes of this experiment suggest that strategic task scheduling may be beneficial in managing tasks with high computational demands, warranting further investigation in this area.

**Table 5: Comparison of DCA vs. FRA Tasks on 140 tasks**

| Task Type | Average. LOC | Average. Time (s) |
|---|---|---|
| DCA | 1,431 | 12.09 |
| FRA | 269,169 | 321.35 |

## 5.4 Evaluating Reusability Designs

To assess reusability, we analyzed average daily calls from real-world production tasks, including security audits and code smell detection. These script-driven tasks, initiated by internal developers across diverse business services, represent authentic usage patterns. This live monitoring demonstrates the system's versatility and the practical effectiveness of our data reuse mechanism in a non-simulated environment.

**Table 6: Comparison of time and memory usage for version code modeling across 8 codebases.**

| Project | LOC | | Time(s) | | Red. | Mem(MB) | | Red. |
|---|---|---|---|---|---|---|---|---|
| | Total | Diff | FRA | IFRA | | FRA | IFRA | |
| rocketmq | 232k | 75 | 107 | 20.93 | 80.38% | 6,417 | 736 | 88.53% |
| dubbo | 263k | 196 | 163.94 | 32.81 | 79.98% | 6,103 | 905 | 85.16% |
| seatunnel | 281k | 320 | 108.8 | 28.64 | 73.66% | 6,821 | 973 | 85.73% |
| storm | 291k | 80 | 176.95 | 21.23 | 88% | 5,718 | 499 | 91.26% |
| iceberg | 488k | 763 | 292.8 | 37.62 | 87.15% | 4,964 | 1,056 | 89.5% |
| pulsar | 598k | 27 | 363.83 | 51.78 | 85.76% | 6,891 | 1,464 | 78.74% |
| hbase | 867k | 625 | 605.31 | 48.19 | 92.03% | 5,535 | 770 | 86.07% |
| iotdb | 1,172k | 89 | 317.8 | 68.26 | 78.51% | 7,812 | 1,618 | 79.27% |
| Avg. | 524k | 471 | 267.01 | 38.68 | 83.18% | 6,282 | 972 | 84.8% |

**Table 7: Comparison of Scheduling Strategies**

| Strategy | Success | Timeouts | Exec. Time (s) |
|---|---|---|---|
| Random | 98.57% | 9 | 10,488 |
| Coordinator | 100% | 0 | 4,006 |

During a week of system operations, we collected task and query statistics (Table 8). The cached COREF facts database had an average reuse factor of 13, indicating each data version was repurposed about 13 times for different static code analysis tasks. In contrast, the average reuse factor for query results was around 1.4 (Table 9), suggesting that while the facts database is highly reusable, specific query demands are more diverse. Notably, general, non-application-specific queries like the Cyclomatic Complexity metric were shared across multiple tasks. This indicates potential in bundling commonly used queries into a "common solution" package to streamline the analysis process and enhance efficiency. Task diversity shows the cached facts database is versatile. Since some queries recur across users, caching common query results (or publishing them via a dedicated interface) can reduce redundant computation.

**Table 8: Average Weekly Statistics of Code Extraction Tasks in Dec 2024.**

| Day | No. of Queries | No. of Extractions | Query/Extraction Ratio |
|---|---|---|---|
| Mon | 119,510 | 7,672 | 15.6 |
| Tue | 126,231 | 8,524 | 14.8 |
| Wed | 132,537 | 9,164 | 14.5 |
| Thu | 133,344 | 9,674 | 13.8 |
| Fri | 110,990 | 11,211 | 9.9 |
| Sat | 17,196 | 1,632 | 10.5 |
| Sun | 9,991 | 768 | 13.0 |
| Avg. | 92,828 | 6,949 | 13.0 |

# 6 Exploring Applications and Use Cases within Ant Group

In this section, we present four use cases of CodeFuse-Query within Ant Group, which has diverse and complex user requirements. To demonstrate the system's potential range of applications, we present an overview of its operational context. The system is currently used by over 30 teams, addressing over 110 distinct daily use cases and 300 annual ones. It processed over 110,000,000 tasks annually, indicating its application to high-volume workloads.

**Table 9: Average Weekly Commit and Query Frequencies for FRA Tasks in Dec 2024.**

| Rev ID | Mon | Tue | Wed | Thu | Fri | Sat | Sun |
|---|---|---|---|---|---|---|---|
| C1 | 3,297 | 3,132 | 2,503 | 2,588 | 2,680 | 352 | 261 |
| C2 | 194 | 408 | 868 | 869 | 437 | 187 | 196 |
| C3 | 194 | 282 | 681 | 868 | 342 | 171 | 187 |
| C4 | 188 | 279 | 661 | 252 | 252 | 153 | 187 |
| C5 | 188 | 221 | 542 | 739 | 216 | 108 | 187 |
| C6 | 187 | 195 | 266 | 252 | 190 | 108 | 95 |
| C7 | 187 | 195 | 252 | 216 | 190 | 95 | 95 |
| C8 | 179 | 194 | 252 | 198 | 190 | 73 | 57 |
| C9 | 177 | 194 | 252 | 198 | 189 | 73 | 57 |
| C10 | 171 | 194 | 201 | 198 | 187 | 68 | 57 |

**(a) Weekly Commits per Revision**

| Script ID | Mon | Tue | Wed | Thu | Fri | Sat | Sun |
|---|---|---|---|---|---|---|---|
| Q1 | 106 | 265 | 148 | 168 | 77 | 40 | 44 |
| Q2 | 86 | 156 | 62 | 64 | 62 | 33 | 17 |
| Q3 | 54 | 51 | 40 | 46 | 51 | 20 | 13 |
| Q4 | 44 | 48 | 37 | 39 | 48 | 12 | 10 |
| Q5 | 38 | 37 | 36 | 37 | 46 | 12 | 10 |
| Q6 | 35 | 34 | 34 | 36 | 36 | 10 | 9 |
| Q7 | 29 | 36 | 34 | 34 | 27 | 10 | 9 |
| Q8 | 29 | 33 | 26 | 26 | 25 | 9 | 8 |
| Q9 | 27 | 30 | 24 | 25 | 25 | 9 | 8 |
| Q10 | 25 | 28 | 24 | 24 | 25 | 9 | 8 |

**(b) Daily Query Frequencies**

### 6.1 Data Preparation for training Large Language Models

Large Language Models (LLMs) are increasingly important in software engineering for tasks such as code generation and analysis [21, 54, 55]. The performance and security of these models heavily rely on the quality and diversity of their training data, which often encompasses billions of lines of code. Handling such vast and diverse code datasets poses significant challenges, particularly with the rapid growth of programming languages and codebases. CodeFuse-Query addresses these challenges by efficiently processing and refining large volumes of code across multiple languages and frameworks. It supports the analysis of massive codebases, identifies syntax errors and security vulnerabilities, and performs necessary transformations to enhance data quality. For instance, CodeFuse-Query can swiftly process multi-terabyte datasets, filter out erroneous code, eliminate common vulnerabilities like SQL injection flaws, and ensure balanced code distribution across various languages to counteract dataset biases. Designed for scalability and adaptability, CodeFuse-Query handles increasing code data volumes and the growing complexity of codebases, meeting current LLM training requirements and anticipating future developments. Using CodeFuse-Query, we have produced a comprehensive dataset comprising 9TB of high-quality code across more than 10 programming languages, which has been utilized to train Ant Group's open-source code LLM, Code Fuse [25].

### 6.2 Metrics-Driven Approach to Enhance Software Engineering Productivity

We aim to cultivate a productivity-centric culture in software engineering by providing developers with actionable insights derived from their code. Leveraging CodeFuse-Query, we deliver precise metrics on code quality, volume, style, and error rates, serving as key indicators for continuous improvement. Our approach operates on two levels: Daily reports offer refined assessments of code changes and highlight errors, supplying targeted improvement suggestions. Seasonal aggregated metrics benchmark performance and encourage healthy competition among developers. With its scalability and precision, CodeFuse-Query efficiently processes large codebases of varying sizes and complexities. Integrating CodeFuse-Query into the development workflow enhances productivity and software quality through continuous improvement.

### 6.3 Change Impact Analysis

"Shift-left" Change Impact Analysis (CIA) [12, 13, 17, 18, 42] is vital in online services risk management, examining code changes to detect potential impacts on services. Integrating CIA into the development cycle enables immediate evaluations post-code submission, requiring rapid analysis—typically within ten minutes. CodeFuse-Query meets this demand and integrates with existing testing environments. The definition of "impact" in CIA is context-dependent; it may represent potential vulnerabilities in security or influence test execution in quality assurance. CodeFuse-Query accommodates diverse impact definitions through customizable Gödel queries, allowing users to tailor CIA to their specific needs—for example, identifying changes affecting certain database tables. By incorporating CIA into development, CodeFuse-Query provides real-time insights, supporting informed decisions on code modifications, security patches, and system stability. This swift analysis enhances the efficiency and effectiveness of the development process.

### 6.4 Ad-hoc Analysis for Large-Scale Codebase Examination

CodeFuse-Query enables comprehensive, ad-hoc analysis for exploring and evaluating large codebases. This capability is crucial for granular, one-time examinations across all repositories. For example, the Quality Assurance team managing Aspect-Oriented Programming (AOP) [41] can use CodeFuse-Query to identify all join point values and affected code segments, ensuring AOP quality. Similarly, Data Compliance teams can conduct holistic scans to identify potential non-compliance with new privacy regulations. CodeFuse-Query facilitates extensive ad-hoc analyses within hours and allows results to be exported for further examination, aiding data-driven decision-making and refinement of architectural practices according to organizational standards.

## 7 Related Work

**Big Data Processing Frameworks:** Big data processing frameworks are pivotal in analyzing large-scale codebases. MapReduce, a programming paradigm designed for processing large-scale data in a distributed environment [24], has influenced numerous systems' design. Hadoop, an open-source framework, facilitates distributed processing of large datasets across hardware clusters [51]. It offers a scalable and fault-tolerant infrastructure for data storage and processing. Spark, another distributed data processing framework, is known for high-speed data analytics, employing an in-memory computing model for extensive data analysis [52]. Higher-level infrastructures, like Hive [49], leverage Hadoop and provide a high-level query language, HiveQL. Hive simplifies data analysis in Hadoop's HDFS by enabling developers to write queries, which are converted into tasks in MapReduce or Spark for efficient evaluation. In our work, we treat code as a specific data form and have developed a DSL, Gödel, for querying large codebases. As a big data processing framework for static code analysis, our approach could gain from big data processing advancements. Future work could involve exploring efficiency enhancements by integrating big data processing algorithms into our technique

**Static Code Analysis Tools:** Static code analysis is a technique that analyzes source code and predicts program behavior without executing the program. To meet the diverse requirements of development, researchers and professionals in both academia and industry have developed various tools with distinct features. For example, Static Application Security Testing (SAST) tools such as Coverity [9] and Klocwork [11] are designed to identify security vulnerabilities in code, which are critical for maintaining software reliability and security. Most of the SAST tools are language-specific, like Clang [1], Infer [19], and Pinpoint [47], which serve as static analyzers for the C family of languages.

In a broader context, tools like SonarQube [4] and PMD [5] offer comprehensive analysis, support multiple languages and provide extensive checks. Besides, there are tools that target specific features such as visualizing code dependencies (CodeScene [8]), enforcing coding standards (StyleCop [3]), or managing complex codebases (CodeRush [7] and JArchitect [10]). These specialized features cater to niche requirements in software development, proving invaluable in specific use cases.

Our work aligns with the emerging concept of a "static analysis ecosystem," which refers to the integration of static analysis within the extensive machinery of large-scale software development. Major technology organizations like Google, Microsoft, Facebook, and Amazon have put this concept into practice by developing their own static analysis tools, including Tricorder [45, 46], CloudBuild [27], and Cloud SAST [22]. Insights from these systems have shed light on the challenges and opportunities of large-scale static analysis. Our research further contributes to this field by proposing a novel solution to enhance the static analysis ecosystem.

**Data-Centric Code Analyzer:** Moor et al. presented CodeQL, an object-oriented query language, for analyzing code [2, 14, 48]. Another effort is Meta's Glean, which was designed to collect and store information about code structure [6]. Our proposed system differs primarily in its design principles compared to these tools. Firstly, our code modeling and extraction are engineered to operate independently of build processes, enhancing the maintainability of CodeFuse-Query (Section 4.2) and achieving better success rates and data extraction performance (Section 5.2). Secondly, CodeFuse-Query's capability to combine two databases for querying enables seamless support for cross-language analysis, allowing CodeFuse-Query to tackle more complex code analysis tasks. Thirdly, CodeFuse-Query is seamlessly integrated into real-world industrial development processes on a daily basis.

# 8 Conclusion

CodeFuse-Query aims to improve large-scale static code analysis by leveraging a data computation approach. Based on the principles of Domain Optimized System Design and Logic Oriented Computation Design, it utilizes resource optimization techniques, task-specific strategies, and Gödel to aim for robustness, scalability, and efficiency in analysis. CodeFuse-Query has been applied to analyze over ten billion lines of code daily, illustrating its capacity for large-scale analysis. The open-sourcing of our computational implementation may contribute to ongoing research and development in the field.